\documentclass[journal]{IEEEtran}

\usepackage{comment}
\usepackage{amsmath}
\usepackage{graphicx}
\usepackage{cite}
\usepackage{url}
\usepackage{color}
\usepackage{tikz}
\usetikzlibrary{arrows.meta, positioning}
\usepackage{algorithm}
\usepackage{algpseudocode}
\usepackage{amssymb}
\usepackage{flushend}
\usepackage{algorithm}
\usepackage{algpseudocode}
\usepackage{booktabs} 
\usepackage{multirow}

\usepackage{graphicx}

\usepackage{tikz}
\usetikzlibrary{arrows.meta,positioning}

\usepackage[caption=false,font=footnotesize]{subfig}
 \usepackage{array}

\ifCLASSINFOpdf
\else

\fi

\allowdisplaybreaks[1]
\renewcommand{\baselinestretch}{0.98}

\begin{document}

\title{Foundation Models Meet Power Grid Topology for Cascade Prediction}

\author{\IEEEauthorblockN{Soumav Biswas and Mia Naeini}

\IEEEauthorblockA{\textit{Electrical Engineering Department, University of South Florida, Tampa, Florida, USA} \\
soumavbiswas@usf.edu, mnaeini@usf.edu}}

\setlength{\textheight}{9.25in}

\maketitle
\thispagestyle{empty}
\begin{abstract}
\textcolor{black}{Accurate early-stage prediction of cascading failure severity is essential for assessing the vulnerability of power transmission networks. However, existing data-driven approaches often rely on network-specific representations, limiting their generalizability across power system topologies. This work investigates the complementary capabilities of Large Language Models (LLMs) and Graph Neural Networks (GNNs) for cascading failure severity prediction. First, standalone LLM predictors are developed using a LoRA-fine-tuned LLaMA-3-8B-Instruct model with varying degrees of topological information. A hybrid Graph-LLM fusion framework is then proposed, integrating a Topology-Aware Graph Attention Transformer (TAGAT) with the LLM through confidence-gated fusion. TAGAT captures structural dependencies among transmission lines, while the LLM extracts representations from early-stage power-flow observations and aggregated cascade descriptors. The frameworks are evaluated on 30,000 simulated cascading failure scenarios for each of the IEEE 118-bus and IEEE 300-bus systems. Using only the first four cascade steps, the hybrid framework achieves weighted F1 scores of 0.94 and 0.92, respectively. In cross-system transfer from IEEE 118-bus to IEEE 300-bus, it achieves 0.72 zero-shot accuracy, improving to 0.83 with limited target-system calibration, outperforming standalone models and conventional baselines. These results show the potential of combining graph-based structural learning with foundation models to improve cascading failure severity prediction and cross-system transferability.}

\end{abstract}

\vspace{-0.2cm}

\vspace{-0.2cm}
\section{Introduction}
Cascading failures are complex, high-impact events that arise as interdependent component failures triggered by initial outages and intensified by shortcomings in control, computation and human error. For decades, these phenomena have been examined from various perspectives to better understand their behavior and develop strategies for prevention, mitigation, and recovery. The techniques used to analyze these phenomena include simulation-based \cite{guo2023review}, physics-driven \cite{zhu2022cascading} and machine learning (ML)-based \cite{pi2018machine} models. 

In particular, ML–based models have recently gained significant attention for cascading failure analysis in power systems to supplement the traditional techniques \cite{sami2024machine}. From classical ML models, including support vector machines (SVMs) and random forests (RFs) \cite{nakas2023online}, to deep neural networks (DNNs) 
and graph-aware models, such as Graph Neural Networks (GNNs) \cite{chadaga2023power}, a range of data-driven approaches has demonstrated strong potential for cascade analyses and prediction {\cite{sami2023machine}. Among these approaches, GNN-based models can utilize network topology and component interactions to capture the complex dependencies underlying cascading failures. However, conventional GNN architectures rely on local neighborhood aggregation and are often coupled to specific network topologies, limiting their system-level reasoning and generalizability across heterogeneous power systems \cite{zhu2025gnns}.

Conversely, Large Language Models (LLMs) have emerged as powerful sequence modeling and reasoning tools for heterogeneous numerical and textual information. For power system applications, LLMs provide a flexible mechanism for representing structured measurements, event descriptors, and system-level properties and
indicators as natural-language prompts \cite{jia2025enhancing}. For cascading failure prediction, LLMs can learn latent relationships between early-stage indicators and eventual cascade severity, providing representations less dependent on fixed graph structures than conventional graph-based models. 

To utilize the complementary strengths of two powerful classes of learning models, namely graph-based architectures and LLMs, this paper adopts a hybrid learning paradigm.  
The main contributions of this work are summarized as follows:

\begin{itemize}
\item A Graph-LLM Fusion framework is proposed for cascading failure severity prediction, integrating topology-aware structural representations learned by Topology-Aware Graph Attention Transformer (TAGAT) with the reasoning and representation capabilities of LLMs through a confidence-gated fusion mechanism. 
\item A comprehensive taxonomy and comparison of cascade severity prediction models is developed across three distinct categories: (i) topology-agnostic, (ii) topology-aware, and (iii) Graph-LLM fusion models. 

\item An extensive set of topology-agnostic baseline models is investigated, including SVM, RF, DNN, Feature Tokenizer Transformer (FT-Transformer), and a standalone LLM variant. 
\item Multiple topology-aware standalone prediction frameworks are evaluated, including TAGAT and LLM-based models enriched with either aggregated topological descriptors or explicit network topology information. 

\item Topological features based on Graph Signal Processing (GSP) are considered to capture global power-flow smoothness and its relationship to uncontrolled islanding. Additional descriptors, including betweenness centrality and bridge properties of initially failed lines, are used to characterize structural vulnerability.

\item Cross-system transferability study evaluates graph-based, LLM-based, and Graph-LLM fusion models trained on the IEEE 118-bus system and tested on the IEEE 300-bus system, assessing zero-shot generalization without parameter adaptation and few-shot calibration using limited target-system data, to address scalability.

\end{itemize}

\vspace{-0.2cm}

\section{Related Works}

Existing data-driven research has focused on classical ML methods and graph-based deep learning frameworks, particularly GNNs. Examples of such studies include cascade size prediction formulated as a classification problem \cite{nakas2023online,sami2023machine,salehpour2022modeling}, component status prediction \cite{bhaila2024cascading,abedi2022svm}, and risk assessment based on contingency precursors \cite{fekri2023vulnerability}. Other applications of data-driven techniques include predicting cascading failure trajectories \cite{he2025vulnerable,liu2020tutorial} and identifying critical components involved in the cascade process \cite{shuvro2019predicting,varbella2023geometric,qi2020utility}.
Moreover, physics-informed graph-based models, with domain constraints or attention mechanisms \cite{yang2022event,zhu2025gnns} have also been proposed for cascade analyses in power systems. GSP methods have also been employed to capture topological relationships and power-flow variations for identifying early indicators of cascading failures \cite{ortega2018graph,sami2024gsp,soumav2025learning}.

 The use of LLMs in power systems analytics is a recent development. Domain-tuned assistants and agent-based systems like GAIA \cite{cheng2024gaia} and GridMind \cite{jin2025gridmind}, utilize LLMs for operational guidance and contingency reasoning. LLM has also been introduced for time-series analysis \cite{matos2025time,lin2025novel} and event forecasting \cite{su2025clb} of real-time data in power systems. 
This approach also has demonstrated notable gains in load forecasting \cite{gao2024general,amjad2025review,li2025zero}, fault diagnosis \cite{zhang2025optimization,jing2024fault}, and event classification \cite{xian2024knowledge}, showing its potential for predictive analytics in power systems.

Recent advances in multimodal learning have demonstrated that combining graph-based models with foundation models enables complementary learning of structural and semantic representations in specific domains \cite{tang2024graphgpt, kang2026togcn,su2026toward}. Although graph-language hybrid frameworks have shown promising results in molecular modeling, knowledge graphs, and recommendation systems, their application to power systems, particularly cascading failure prediction, remains largely unexplored. Motivated by these advances, this work integrates topology-aware graph representation learning with LLM reasoning capabilities for early cascade severity prediction.

Moreover, cross-system transferability remains a key challenge for many ML models, particularly those developed for power systems with different network topologies. Power networks may differ significantly in bus count, line count, graph density, and operating conditions. Recent studies on GNNs for power system applications have shown that topology changes and distribution shifts can strongly affect model generalization, motivating topology-aware but transferable learning strategies \cite{zhu2025gnns,deJong2025generalizable,wu2025universal}. Domain adaptation techniques, such as correlation alignment, can reduce distribution mismatch between source and target systems by aligning second-order feature statistics \cite{sun2016deep}. In addition, few-shot calibration and source-to-target adaptation methods provide practical mechanisms for adapting a trained model using limited target-domain information \cite{liang2020shot,kumar2020transcal}. These ideas are especially relevant for hybrid graph-language models, where topology-aware representations may improve in-distribution prediction but may also require adaptation when deployed on unseen network topologies.

To the best of our knowledge, existing studies have not systematically compared topology-aware graph representation learning, topology-agnostic LLMs, and their hybrid fusion for cascading failure analysis in power systems. This paper addresses these shortcomings by systematically evaluating these formulations in terms of in-distribution predictive accuracy, zero-shot cross-system transferability, and few-shot adaptation under a limited target-label budget.

\vspace{-0.2cm}

\section{Cascade Simulation, Data Generation, and Feature Definition}
This section describes the cascading failure model used in the simulation and the methodology used for generating data description from the simulated cascading failure events.

\subsubsection{\textbf{Cascading Failures Model}}
The cascading failure events are modeled using a line-overloading mechanism implemented in MATPOWER, where DC power flow is recomputed after each outage to determine post-contingency line loading \cite{wu2020influence}. The cascade continues until no additional overloaded lines are detected. 
Two systems, specifically, the IEEE-118 and IEEE-300 test systems have been used for generating the cascade events and for the performance evaluation of the models. The IEEE 118-bus (118 buses, 186 lines, 54 generators, 91 loads) and IEEE 300-bus (300 buses, 411 lines, 69 generators, 200 loads) systems serve as standard benchmark models.

It is assumed that two initial failures trigger the cascading process, after which subsequent steps involve a single failure at a time caused by line overloading. This simplification states that time can be divided into sufficiently small intervals, allowing multiple near-simultaneous failures to be treated as sequential events occurring in rapid succession.
 
The cascade scenarios are grouped by the severity of the outage: {\it Small} (1 to 10\%), {\it Medium} (10 to 30\%), {\it large} (30 to 60\%), and {\it Very large} ($>$60\%) number of transmission lines failed. Each dataset contains 30,000 cascading failure scenarios, with 3,466 small, 13,303 medium, 10,108 large, and 3,123 very large cascades for IEEE-118, and 2,964, 12,443, 13,206, and 1,387, respectively, for IEEE-300. The objective is to predict the severity label \(y\in\{0,1,2,3\}\), representing small, medium, large, and very large cascades, respectively.

\subsubsection{\textbf{System Model as a Graph}}

The power transmission network is represented as a line graph \(G=(V,E)\), where each node
\(v_i\in V\) represents transmission line \(i\), and each edge \((v_i,v_j)\in E\) indicates that the two transmission lines share a common bus.
Unlike the conventional bus graph, the line graph captures the connectivity among transmission lines through which cascading failures propagate. For each benchmark system,
the line graph is constructed once and remains fixed across cascading failure
scenarios, while node attributes evolve with the cascade.

\subsubsection{\textbf{Node Features}}

For the first \(T\) observed cascade steps, the feature vector associated with transmission line \(i\) is defined as \(\mathbf{x}_i=(p_i^{(1)},p_i^{(2)},\ldots,p_i^{(T)}),\)  where
\(p_i^{(t)}\) denotes the active power flow through transmission line \(i\) after cascade step \(t\). These node features form the graph signal matrix \(
\mathbf{X}\in\mathbb{R}^{L\times T},
\) where \(L\) denotes the total number of transmission lines.

\subsubsection{\textbf{Global Features}}

In addition to localized node-level features, the proposed framework incorporates a set of global cascade descriptors that summarize the overall operating conditions and cascade behaviors, providing complementary information beyond local graph neighborhoods.

{\bf \emph{GSP-based Global Features:}} Following the GSP framework proposed in \cite{soumav2025learning}, this work defines four GSP-derived features for each cascade case to capture the global characteristics of the cascading event. In the following paragraphs, first the key GSP concepts used to define these features are reviewed and then the four features are presented.

The graph signal $\mathbf{X}$ introduced earlier represents the power-flow values on the transmission lines, which are modeled as nodes in the line graph. Variations in the graph signal reflect the degree of unevenness in power-flow distribution across electrically adjacent transmission lines. To quantify this variation, {\it graph signal smoothness} is used as a system-level indicator of power-flow variation over the line graph. At each cascade step $t$, the local smoothness, $s_i^{(t)}$, of transmission line $i$ quantifies the deviation of its power flow from the average power flow of its neighboring lines,
\(s_i^{(t)}=\left|p_i^{(t)}-\frac{1}{|\mathcal{N}(i)|}\sum_{j\in\mathcal{N}(i)}p_j^{(t)}\right|,\) where \(\mathcal{N}(i)\) denotes the set of neighboring lines connected to line \(i\) in the line graph. The global smoothness is then obtained by aggregating the local smoothness values over all transmission lines,
\(S^{(t)}_{\mathrm{global}}=\sum_{i=1}^{L}s_i^{(t)}\). A larger value of \(S^{(t)}_{\mathrm{global}}\) indicates greater spatial variation in power flows across adjacent transmission lines, which may reflect increasing system stress during the cascade.

The analysis presented in \cite{soumav2025learning} showed that the first local maximum of the global smoothness trajectory often occurs before the first uncontrolled islanding event, indicating that the system experiences a stress peak before losing network connectivity, motivating the use of graph smoothness-based indicators as early descriptors of cascade severity. Accordingly, four GSP-derived features are extracted for each cascading failure scenario:
\begin{itemize}
    \item {\bf \textit{GF1:}} The cascade step at which the first local maximum of $S^{(t)}_{\mathrm{global}}$ occurs. 
    \item {\bf \textit{GF2:}} The cascade step at which the first uncontrolled islanding event occurs. 
    \item {\bf \textit{GF3:}} The step distance between these two events (between GF1 and GF2).
    \item  {\bf \textit{GF4:}}  Amplitude of the first local maximum of $S^{(t)}_{\mathrm{global}}$.
\end{itemize}

{\bf \emph{non-GSP-based Global Features:}} In addition to the GSP-derived features, sets of global descriptors are extracted from the initial failures and the early-stage cascade trajectory. 

Specifically, for each of the two initial line failures, three static structural descriptors are computed from the fixed line-graph topology. The first is the bridge line indicator meaning whether the failed line is a bridge, i.e., whether its removal disconnects the line graph. Second one is
the line's betweenness centrality, defined as the fraction of shortest paths between pairs of other lines that pass through the line. And the third one is 
line's degree, representing the number of lines adjacent to it in the graph. 
These features are referred to as:
\begin{itemize}
  \item \textbf{\textit{GF5:}} The structural descriptors (bridge indicator, degree and betweenness centrality) of each of the two initially failed lines.
\end{itemize}

\textcolor{black}{ To further characterize the system state, root-mean-square (RMS) power-flow indicators are computed at each observed cascade step to summarize the global loading level. At cascade step $t$, the RMS power-flow indicator is defined as
\(R^{(t)}=\sqrt{\frac{1}{L}\sum_{i=1}^{L}\left(p_i^{(t)}\right)^2}\).}

\textcolor{black}{Evaluating this expression at each of the $T$ observed cascade steps yields the sequence of scalar values $R^{(1)}, \ldots,
R^{(T)}$.} 
These features are referred to as:
\begin{itemize}
  \item \textbf{\textit{GF6:}} The RMS power-flow indicators $R^{(1)},\ldots,R^{(T)}$.
\end{itemize}

\textcolor{black}The mean graph distance between pairs of failed lines within the observed cascade window of size $T$ is also considered. The mean graph distance is the average shortest path distance in hops along the line-connectivity graph between the failed lines,} denoted as:
\begin{itemize}
  \item \textbf{\textit{GF7:}} The mean graph distance among the lines that have failed within the observed window of size $T$.
  \end{itemize}

Finally, two network-level descriptors characterize the system itself and are identical for every scenario on a given network. These include 
the average node degree, $\bar{d}=\frac{1}{L}\sum_{i=1}^{L}|\mathcal{N}(i)|$, and the graph density, $\rho=\frac{2|E|}{L(L-1)}$, where $\mathcal{N}(i)$ is the set of lines adjacent to line $i$ \cite{barabasi2016network}. 
The density $\rho$ is the fraction of line pairs that share a bus. These features are referred to as:
\begin{itemize}
\item
  \textbf{\textit{GF8:}} The network-level descriptors including the average node degree, $\bar{d}$, and graph density of the line graph, $\rho$.
\end{itemize}

\vspace{-0.2cm}

\section{Cascade Prediction Framework and Model Architecture}

This section reviews and presents the class-based cascade size prediction frameworks considered in this study. The models are organized into three categories according to the extent to which they incorporate structural information about the system's graph:
 
{\bf \emph{(A) Topology-agnostic baseline models,}} which operate solely on tabular or sequential feature representations without explicitly incorporating the network structure. The models in this category include SVM,  RF, DNN, FT-Transformer, and a standalone LLM model.
{\bf \emph{(B) Topology-aware standalone models,}} which incorporate explicit line-graph structure or structural descriptors as part of their input. This category includes the TAGAT and two standalone LLM models with topological information.
{\bf \emph{(C) Graph–LLM fusion model,}} which combines both representations through a confidence-gated fusion module.

\vspace{-0.2cm}
\subsection{\textbf{Topology-Agnostic Baseline Models}}

The following five models serve as topology agnostic baselines. Each is trained and evaluated using the same train/validation/test
splits as the other presented models, without access to the line-graph adjacency or any structural descriptor of the transmission network.

\subsubsection{\textbf{Support Vector Machine (SVM)}}
An SVM classifier with a radial basis function (RBF) kernel is used as a non-parametric baseline. The RBF kernel
enables the model to learn nonlinear decision boundaries between cascade severity classes. 

\subsubsection{\textbf{Random Forest (RF)}}
An RF classifier is used as an ensemble tree-based baseline. By aggregating predictions across a large number of decorrelated decision trees trained on bootstrap-resampled subsets of the data, the RF model provides a robust, low-variance baseline that does not require explicit topological information.

\subsubsection{\textbf{Deep Neural Network (DNN)}}
The DNN baseline consists of 
\textcolor{black}{an input layer whose size equals the dimension of the feature vector, depending on the size of the system and the feature set, followed by four fully connected hidden layers of widths 128, 128, 256, and 256, using ReLU activations and a dropout rate of 20\% for regularization.} The network is trained with the Adam optimizer at a learning rate of $1\times10^{-4}$, using categorical cross-entropy as the loss function, for 120 epochs with a batch
size of 64 and a validation split of 20\%.

\subsubsection{\textbf{Feature Tokenizer Transformer (FT-Transformer)}}
As a more advanced topology-agnostic baseline for tabular data, an FT-Transformer is also implemented. Unlike recurrent or graph-based architectures, the FT-Transformer employs multi-head self-attention directly on
feature tokens, allowing global interactions among input features without relying on explicit graph topology modeling. 

\subsubsection{\textbf{ No Topology Information LLM (NTI-LLM)}} 

\begin{figure}
    \centering    \includegraphics[width=3.4in]{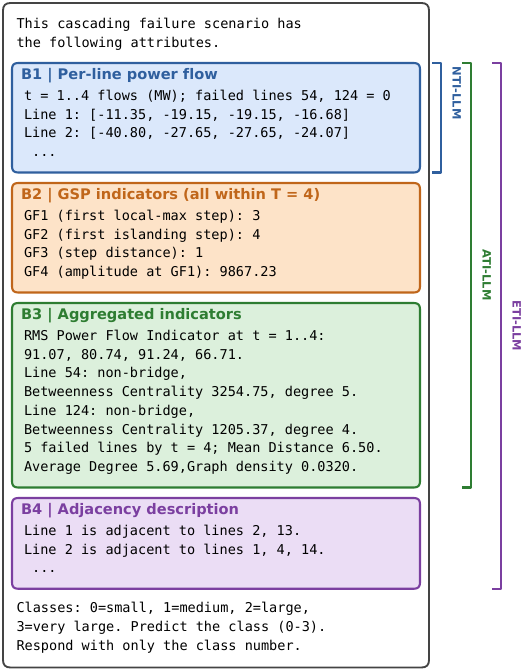}
    \caption{\footnotesize{Example testing prompt for a scenario in which all four GSP indicators are observed within the first $T=4$ steps. The brackets mark the blocks used by each variant: (a) NTI-LLM uses B1 (per-line power flow); (b) ATI-LLM uses B1-B3 (adds the GSP and Non-GSP Global indicators); (c) ETI-LLM uses B1-B4 (adds the line-adjacency description).
    }}
    \label{prompt}
    \vspace{-0.2cm}
\end{figure}

The first variant of standalone LLM model developed in this study predicts cascade severity directly from a structured natural-language prompt. It does not use a graph encoder, line-graph representation, or any explicit topological information. \textcolor{black}{Instead, the model uses only the per-line active power flow, i.e., the vector $\mathbf{x}_i=(p_i^{(1)},\ldots,p_i^{(T)})$ of every transmission line $i=1, \ldots,L$.} Figure~\ref{prompt} depicts the structure of the prompt, which consists of information blocks, where each block is relevant to a specific variant of the LLM model in this paper.
This variant of LLM uses a single block, marked as B1, which contains the per-line active power flow $p_i^{(t)}$ for $t=1,\ldots,T$. The rest of the information blocks will be introduced along with the other LLM variants. During supervised fine-tuning in training, the severity label is appended to the prompt as the expected output; during testing, it is omitted and the LLM predicts the severity class.

The open-source Meta-LLaMA-3-8B-Instruct model serves as the LLM backbone. Its decoder architecture incorporates multi-head self-attention, rotary position embeddings, and automatic tokenization, enabling cascade severity prediction to be formulated as a conditional generation task.

Low-Rank Adaptation (LoRA) offers parameter-efficient finetuning by injecting small-rank decomposition matrices into the model's attention projections while also keeping the original weights of the model unchanged \cite{hu2022lora}.
LoRA adapters are applied to the attention projection matrices \(\mathbf{w}_{q}, \mathbf{w}_{k}, \mathbf{w}_{v}, \mathbf{w}_{o}\), 
corresponding to the query, key, value, and output transformations in each attention layer, while the other parameters were frozen. 

LoRA modifies the existing weight matrix $\mathbb{W}$ using a low-rank update: $\mathbf{W'} = \mathbf{W} + \mathbf{B}\mathbf{A}$, where $\mathbf{A} \in \mathbb{R}^{r \times d}$,  and $\mathbf{B} \in \mathbb{R}^{d \times r},$ \textcolor{black}{ where,  $d$ is the hidden dimension of the attention projection matrix
being adapted and $r$ is the LoRA rank.}

Because $r \ll d$, the number of trainable parameters is reduced considerably to make the model size compatible with the available computational resources.
The objective function of the fine-tuning is the standard causal language modeling (CLM)
loss. 

\textcolor{black}{ Given a scenario prompt $s$ and its associated label token $y$, the
loss is $\mathcal{L}= - \log P_{\theta}(y \mid s),$ which is computed only over the final label token.}

Fine-tuning and training are performed using the HuggingFace Trainer, which enables automatic checkpointing at each epoch and loads the best model based on lowest validation loss. All prompt tokens
are masked with a loss value of $-100$, ensuring that only the prediction of
the cascade class contributes to the gradient updates. Table \ref{config} presents the model initialization and the associated parameters tuned for optimizing the predictor. \textcolor{black}{Due to GPU memory constraints associated with fine-tuning of the 8B-parameter model, the per-device batch size was set to 1. Gradient accumulation with 8 steps was employed to maintain an effective batch size of 8, ensuring stable optimization while preserving computational feasibility.} Rather than relying on autoregressive label generation, the proposed implementation extracts both the semantic embedding and class-probability estimate from a single forward pass through the fine-tuned LLM. The final-layer hidden states are mean-pooled over the prompt tokens to obtain the semantic embedding \(\mathbf{h}_L\in\mathbb{R}^{4096}\), \textcolor{black}{where 4096 is the hidden state dimension of the Meta-LLaMA-3-8B-Instruct backbone model.} 
In parallel, a four-dimensional class-probability vector \(\boldsymbol{\pi}_L\) 
is obtained from the next-token logits by applying softmax over the token IDs corresponding to the four class labels.

\begin{table}[h!]
\centering
\caption{\footnotesize {\textcolor{black}{Model Parameters and Configuration for the Standalone LLM-based Prediction Models}}} 
\resizebox{\columnwidth}{!}{%
\begin{tabular}{|l|l|l|}

\hline

\textbf{Category}       & \textbf{Parameter}        & \textbf{Value} \\ \hline

\multirow{5}{*}{\textbf{Model}} 
    & Base model           & Meta-LLaMA-3-8B-Instruct \\ \cline{2-3}
    & Number of Attention Heads & 32 \\ \cline{2-3}
    & Train/Val/Test Split & 70\% / 10\% / 20\% \\ \cline{2-3}
    & Epochs               & 3 \\ \cline{2-3}
    & Max sequence length  & 768 \\ \hline

\multirow{5}{*}{\textbf{Optimization}} 
    & Optimizer            & AdamW \\ \cline{2-3}
    & Per-device batch size & 1 \\ \cline{2-3}
    & Gradient accumulation & 8 (effective batch = 8) \\ \cline{2-3}
    & Learning rate        & \(1 \times 10^{-4}\) \\ \cline{2-3}
    & Weight decay         & 0.02 \\ \hline

\multirow{4}{*}{\textbf{LoRA}} 
    & Rank (\(r\))         & 16 \\ \cline{2-3}
    & Alpha                & 32 \\ \cline{2-3}
    & Dropout              & 0.03 \\ \cline{2-3}
    & Target modules       & \({w_q},\, {w_k},\, {w_v},\, {w_o}\) \\ \hline

\textbf{Precision}      & Mixed Precision     & bf16 \\ \hline

\textbf{Hardware}       & GPU                 & NVIDIA A100 \\ \hline
\end{tabular}
}
\label{config}
\end{table}

\vspace{-0.2cm}

\subsection{\textbf{Topology-Aware Standalone Models}}

The following three models incorporate structural information about the transmission network, either through the explicit line graph or through structural descriptors derived from it. 

\begin{figure*}[t]
    \centering
    \includegraphics[width=\textwidth]{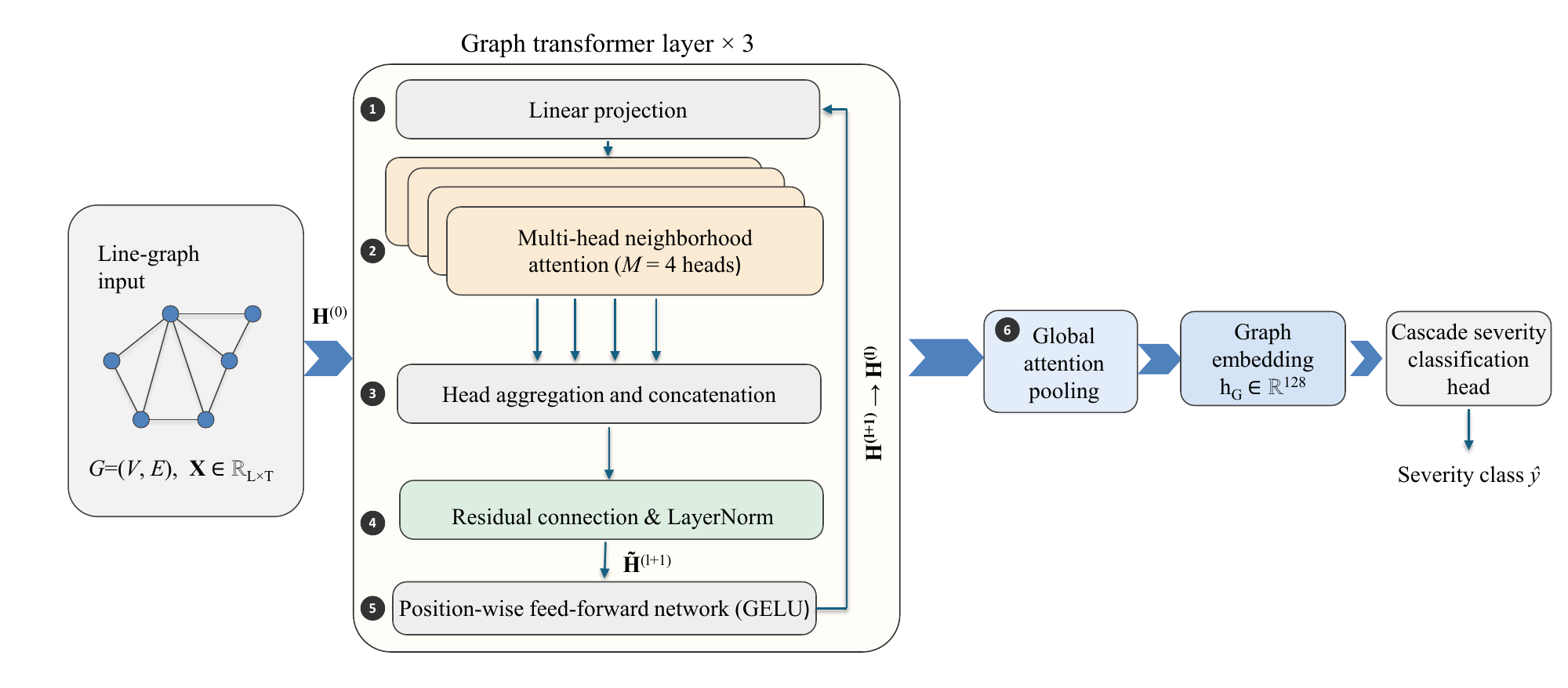}
   
    \caption{Architecture of the standalone TAGAT model. Three stacked graph transformer layers perform multi-head
neighborhood attention, residual normalization, and feed-forward transformations. Global attention pooling produces the graph embedding $h_G \in \mathbb{R}^{128}$, used for cascade severity classification.}
    \label{fig:tagat_architecture}
\end{figure*}

\subsubsection{\textbf{Topology-Aware Graph Attention Transformer (TAGAT)}}

The proposed TAGAT encoder consists of three stacked graph transformer layers followed by a global pooling stage, as shown in
Fig.~\ref{fig:tagat_architecture}. Each layer is built from five blocks  applied in sequence to the per-line hidden representation $\mathbf{H^{(l)}}\in\mathbb{R}^{L\times d}$, where $L$ is the
number of transmission lines, $d$ is the hidden dimension, and the rows $h_i^{(l)}$ are the hidden vectors of the individual lines. The output
of block (\textit{5}) is the input of the next layer, and block (\textit{6}) is applied once, after the third layer.

\textit{{\bf Block 1}-Linear projection of query, key, and value vectors:}  Each layer first projects the node features into a hidden representation before computing query, key, and value vectors,  \(
\mathbf{Q}=\mathbf{H}^{(l)}\mathbf{W}_Q, \mathbf{K}= \mathbf{H}^{(l)}\mathbf{W}_K, \mathbf{V}= \mathbf{H}^{(l)}\mathbf{W}_V,
\) \textcolor{black}{where 
$\mathbf{H}^{(l)} \in \mathbb{R}^{L\times d}$ denotes the hidden representation at the $l$-th layer, whose rows are the per-line hidden vectors $h_i^{(l)}$, and $\mathbf{H}^{(l)}$ is the full set of
node vectors $h_i^{(l)}$. $\mathbf{W}_Q$, $\mathbf{W}_K$, and $\mathbf{W}_V$ are learnable projection matrices. Moreover, $q_i = h_i^{(l)}\mathbf{W}_Q$,
$k_i = h_i^{(l)}\mathbf{W}_K$, and $v_i = h_i^{(l)}\mathbf{W}_V$ denote the row-wise query, key, and value vectors for line $i$, in $\mathbf{Q}$, $\mathbf{K}$, and
$\mathbf{V}$, respectively.} 

\textit{{\bf Block 2}- Multi-head neighborhood attention:} For every transmission line, attention coefficients are computed over its neighboring transmission lines using scaled dot-product attention, as \( \alpha_{ij}={\exp\left(\frac{\mathbf{q}_i^T\mathbf{k}_j}{\sqrt{d}}\right)}/{\sum\limits_{k\in\mathcal N(i)}\exp\left(\frac{\mathbf{q}_i^T\mathbf{k}_k}{\sqrt{d}}\right)},\) where \(\mathcal N(i)\) denotes the neighboring transmission lines connected to line \(i\), and \(d\) is the hidden embedding dimension. The attention coefficients quantify the influence of neighboring transmission lines, while four independent attention heads capture different neighborhood interaction patterns during cascade evolution.

\textit{{\bf Block 3}- Aggregation and head concatenation:} The hidden representation of each node is then updated through weighted aggregation within each head, \(\mathbf{a}_{i,m}^{(l)}=\sum_{j\in\mathcal N(i)}\alpha_{ij}\mathbf{W}_O
\mathbf{v}_j,\) \textcolor{black}{where \(\mathbf{W}_O\) is a learnable output projection matrix and $v_j$ is the value vector of neighboring line $j$. The outputs of the heads are then concatenated, \(\mathbf{a}_i^{(l)}=\big\Vert_{m=1}^{M}\mathbf{a}_{i,m}^{(l)},\) where \(M=4\) denotes the number of attention heads and
\(\Vert\) represents feature concatenation. Here, $\mathbf{a}_{i,m}^{(l)}$ is the attention output of line $i$ computed independently by the $m$-th head, using that head's own attention weights and projection matrices, and $\mathbf{a}_i^{(l)}$ is the resulting multi-head output of line $i$.}

\textit{{\bf Block 4}- Residual connection and layer normalization:}
Following each graph attention layer, residual connections and layer normalization are applied to stabilize optimization and facilitate gradient propagation through deeper architectures, 

\(\widetilde{\mathbf{H}}^{(l+1)}=\mathrm{LayerNorm}\left(\mathbf{H}^{(l)}+\mathrm{TAGAT}(\mathbf{H}^{(l)})\right),\)  
where $\mathrm{TAGAT}(\mathbf{H}^{(l)})$ denotes the matrix obtained by stacking the per-node multi-head outputs $\mathbf{a}_i^{(l)}$ of Block 3 row-wise, while both terms of the sum are computed at layer $l$, and $\widetilde{\mathbf{H}}^{(l+1)}$ denotes this post-attention, pre-FFN state.

\textit{{\bf Block 5}- Position-wise Feed-Forward Network:} The resulting node representations are subsequently refined through a position-wise feed-forward network consisting of two fully connected layers with GELU activation, \(\mathrm{FFN}(\mathbf{h})=\mathbf{W}_2\,\mathrm{GELU}(\mathbf{W}_1
\mathbf{h}),\) 
\textcolor{black}{applied identically to each node's hidden vector $h$, where $h = \tilde{h}_i^{(l+1)}$, the $i$-th row of $\widetilde{\mathbf{H}}^{(l+1)}$ for each line in turn. This is the post-attention, pre-FFN vector of line $i$, not the layer input $h_i^{(l)}$ or the attention outputs $a_{i,m}^{(l)}$ and $a_i^{(l)}$. $\mathbf{W}_1$ and $\mathbf{W}_2$ are the feed-forward layer's learnable weight matrices. }\textcolor{black}{The hidden dimension of the feed-forward network is set to 256. Dropout with probability 0.1 is applied after both the attention and feed-forward blocks to reduce overfitting. Applying the FFN row-wise to every line's hidden vector yields the layer's final output $\mathbf{H}^{(l+1)}$, whose rows are $h_i^{(l+1)}=\mathrm{FFN}(\tilde{h}_i^{(l+1)})$. This becomes the input $\mathbf{H}^{(l)}$ of the next graph transformer layer; continuing up to the third layer before global pooling.}

\textit{{\bf Block 6}- Global attention pooling:} After the third and final TAGAT layer, the node embeddings are aggregated using global attention pooling to obtain a graph-level representation, \(h_G=\sum_{i=1}^{N}\beta_i h_i,\)
where \(\beta_i=\exp(g(h_i))/\sum_j \exp(g(h_j))\) represents the learned attention weight assigned to transmission line \(i\), and \(g(\cdot)\) is a learnable scoring network \cite{shi2021masked}. 
The final graph embedding \(\mathbf{h}_G\in\mathbb{R}^{128}\) encodes the structural characteristics of the cascading failures. The architecture of Standalone-TAGAT model is shown in Fig. \ref{fig:tagat_architecture}.

During training, topology augmentation is applied by sampling random induced subgraphs from the full line graph. For each training instance, a fraction of the line nodes is sampled uniformly between 0.5 and 1.0, with a minimum of eight nodes. The adjacency matrix and node features are used for training, while validation, testing, and target-system evaluation use the full original line graph. This augmentation prevents the graph encoder from overfitting to a fixed network size. To improve cross-system representation alignment, Correlation Alignment (CORAL) loss is applied during training between source-system and unlabeled target-system graph embeddings by penalizing differences between their covariance matrices \cite{sun2016deep}. This alignment requires no target-system labels and is not updated during subsequent target-system evaluation.

\subsubsection{\textbf{Aggregated Topology-Informed LLM (ATI-LLM)}}

\textcolor{black}{While the LLM formulation described in Section IV.A operates only on per-line power flow data, this topology-aware variant of the standalone LLM classifier incorporates some aggregated topological information in the form of natural language for the classification. This model shares the same LLaMA-3-8B-Instruct backbone and LoRA fine-tuning procedure detailed in Section IV.A; the only distinction from the topology-agnostic formulation lies in two additional information blocks appended to the prompt after information block B1 in Fig. \ref{prompt}. These information blocks include (B2) the GSP-based features including GF1 to GF4, and (B3) the non-GSP-based global features including GF5 to GF8, as explained in Section~III.}

\textcolor{black}{To prevent information leakage beyond the first $T$ observed steps, GSP features in block (B2) are conditionally included based on event occurrence. If both the first local maximum and uncontrolled islanding occur, GF1--GF4 are included; if only the local maximum occurs, GF1 and GF4 are included; and if only islanding occurs, GF2 is included. If neither occurs, no GSP feature values are provided. Unavailable indicators are explicitly identified in the prompt.}

\subsubsection{\textbf{Explicit Topology-Informed LLM (ETI-LLM)}}  
\textcolor{black}{This topology-aware variant of the standalone LLM classifier incorporates explicit knowledge of the network topology in the form of adjacency matrix information in natural language classification. This model shares the same LLaMA-3-8B-Instruct backbone and LoRA fine-tuning procedure detailed in the LLM presented previously;
the only distinction from the ATI-LLM lies in a fourth information block (B4), which is a natural-language description of the line-graph adjacency that lists, for each line, the lines adjacent to it (see Fig.~\ref{prompt}). The information blocks B1--B3 and the conditional disclosure of the GSP features in (B2) are identical to those of the previous variants, so the same reduced (B2) block, as discussed earlier, is used whenever an event is not observed within the window.}

\begin{figure*}[!t]
    \centering
    \includegraphics[width=\textwidth]{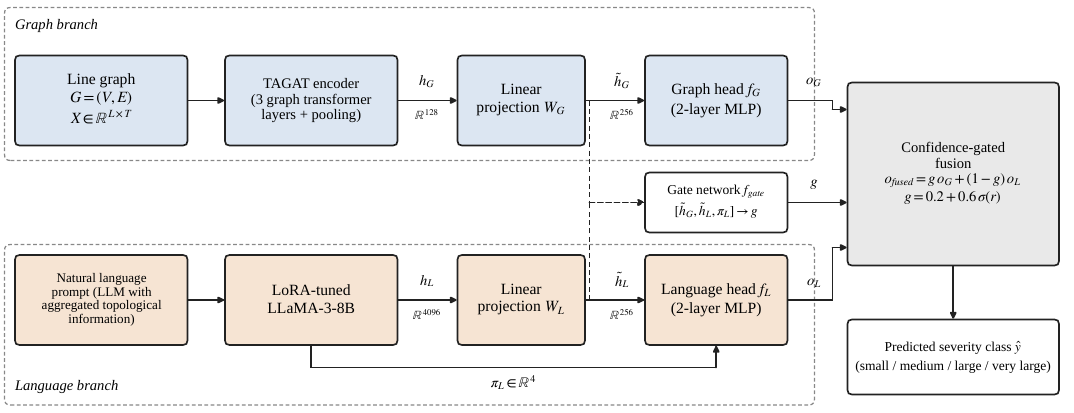}
    
     \caption{Architecture of the proposed Graph-LLM Fusion framework. The TAGAT and ATI-LLM branches generate graph and language representations, respectively, which are projected into a shared latent space. Separate classification heads produce logits that are combined through a confidence-gated fusion mechanism for severity prediction.}
    \label{fig:hybrid_framework}
\end{figure*}

\vspace{-0.2cm}
\subsection{\textbf{Graph-LLM Fusion Network}} 
The hybrid Graph-LLM Fusion framework includes two branches: a TAGAT branch and an LLM branch with aggregated topological features and unifies the two within a single end-to-end classifier, as illustrated in Fig.~\ref{fig:hybrid_framework}. Since the TAGAT branch uses explicit topological information, the chosen LLM is the ATI-LLM. 
Note that the LLM branch receives the reduced (B2) information block shown in Fig.~\ref{prompt} whenever a GSP event is not observed within the window of size $T$, while the TAGAT branch uses no GSP features and is unaffected.  
\textcolor{black}{The framework employs decision-level fusion, where class logits from both branches are combined through a learned confidence gate, which enables scenario-specific weighting of graph-based and language-based predictions.} 
\textcolor{black}{The remainder of this subsection first presents the confidence-gated fusion module, which turns the graph and language embeddings into a single prediction, and then the gate regularization terms and the composite training objective used to train it.}

\subsubsection{Confidence-Gated Graph-Language Feature Fusion}

The proposed fusion module combines graph-based and language-based class logits through a learned confidence gate, enabling scenario-specific weighting of both branches. 

Let \(\mathbf{h}_G\in\mathbb{R}^{128}\) denote the graph-level embedding generated by the TAGAT encoder after graph attention pooling, and let \(\mathbf{h}_L\in\mathbb{R}^{4096}\) denote the semantic embedding extracted from the final hidden states of the utilized LLM model. In addition to the semantic embedding, the LLM branch also produces a four-dimensional class-probability vector \(\boldsymbol{\pi}_L\in\mathbb{R}^{4}\), 

Since the graph and language embeddings originate from different feature spaces and have different dimensionalities, they are first projected into a common latent space using independent learnable projection layers,\(\tilde{\mathbf{h}}_G=\mathbf{W}_G\mathbf{h}_G+\mathbf{b}_G,\) \(\tilde{\mathbf{h}}_L=\mathbf{W}_L\mathbf{h}_L+\mathbf{b}_L,\)
where \(\mathbf{W}_G\in\mathbb{R}^{256\times128}\) and \(\mathbf{W}_L\in\mathbb{R}^{256\times4096}\) are trainable projection matrices. The projected graph embedding \(\tilde{\mathbf{h}}_G\) is passed through a graph-specific classification head to produce the graph-branch logits, \(\mathbf{o}_G=f_G(\tilde{\mathbf{h}}_G).\) Similarly, the language-specific classification head receives both the projected language embedding and the LLM class-probability vector. The language-branch logits are computed as \(\mathbf{o}_L=f_L\left(\mathrm{concat}\left(\tilde{\mathbf{h}}_L,\boldsymbol{\pi}_L\right)\right),\)
where \(f_G(\cdot)\) and \(f_L(\cdot)\) denote two-layer multilayer perceptrons with GELU activation and dropout. Including \(\boldsymbol{\pi}_L\) allows the language head to use both the semantic representation learned from the prompt and the LLM's own class-probability estimate.

To adaptively combine the two branches, a separate gate network is introduced. The gate network takes the projected graph embedding, the projected language embedding, and the LLM probability vector as input, \(\mathbf{u}=\mathrm{concat}\left(\tilde{\mathbf{h}}_G,\tilde{\mathbf{h}}_L,\boldsymbol{\pi}_L\right),\) and produces a raw scalar gate logit,
\(r=f_{\mathrm{gate}}(\mathbf{u}).\) The raw gate value is passed through a sigmoid function and then affinely rescaled to prevent the fusion module from collapsing completely onto either branch, \(g=0.2+0.6\sigma(r).\)
Thus, the gate is bounded as \(0.2\leq g\leq0.8.\) The final fused logits are obtained through a confidence-weighted combination of the graph and language logits, \(\mathbf{o}_{\mathrm{fused}}=g\mathbf{o}_G+(1-g)\mathbf{o}_L.\)
When \(g\) is close to its upper bound, the model assigns greater confidence to the TAGAT branch. When \(g\) is close to its lower bound, the model relies more heavily on the LLM branch. Since the gate is bounded, both branches always contribute to the fused decision, preventing the hybrid framework from degenerating into a single-branch classifier. The fused logits are converted into class probabilities through the softmax function, \(P(y=c)=\frac{\exp(o_{\mathrm{fused},c})}{\sum_{k=1}^{C}\exp(o_{\mathrm{fused},k})},\)
 The final predicted class is obtained as \(\hat{y}=\arg\max_c P(y=c).\)

\subsubsection{Gate Regularization and Composite Training Objective}

Bounding $g$ to $[0.2,0.8]$ prevents either branch from being fully excluded, but not from saturating near a bound for most scenarios. Two lightweight penalties discourage this: an entropy term $\mathcal{L}_{H}=-\frac{1}{N}\sum_i[\hat{g}_i\log\hat{g}_i+(1-\hat{g}_i)\log(1-\hat{g}_i)]$, with $\hat{g}_i=(g_i-0.2)/0.6$, which is maximized (via a subtracted loss term) to keep per-example gates away from either bound; and a batch-balance term $\mathcal{L}_{\mathrm{bal}}=(\bar g-0.5)^2$, with $\bar g=\frac1N\sum_i g_i$, which keeps the batch-average gate centered so the module does not systematically favor one branch. A second CORAL term, $\mathcal{L}_{\mathrm{CORAL}}^{\mathrm{fusion}}$, is also applied to the concatenated projected embeddings $\tilde{\mathbf{h}}=\mathrm{concat}(\tilde{\mathbf{h}}_G,\tilde{\mathbf{h}}_L)$ using source features, complementing the TAGAT-level CORAL term of Section~IV-B since the projection and classification heads sit downstream of the graph encoder and can drift out of alignment independently.

\vspace{-0.2cm}
\section{Performance Evaluation}
This section reports the empirical comparison of all model families introduced in Section IV.

\begin{table}[t]
\centering
\caption{Cascade Severity Classification Performance on IEEE-118  bus system using T=4}
\label{class118}
\small
\setlength{\tabcolsep}{3pt}
\begin{tabular}{>{\raggedright\arraybackslash}p{1.85cm}>{\raggedright\arraybackslash}p{2.7cm}ccc}
\toprule
Category & Model & F1 & Precision & Recall \\
\midrule
\multirow{3}{*}{ML}
& SVM-RBF          & 0.85 & 0.86 & 0.85 \\
& Random Forest    & 0.94 & 0.93 & 0.95 \\
& DNN              & 0.86 & 0.87 & 0.86 \\
\midrule
\multirow{1}{*}{Transformer}
& FT-Transformer   & 0.88 & 0.89 & 0.87 \\
\midrule
\multirow{3}{*}{LLM}
& NTI-LLM & 0.90 & 0.91 & 0.89 \\
& ATI-LLM & 0.93 & 0.93 & 0.92 \\
& ETI-LLM & 0.93 & 0.93 & 0.93 \\
\midrule
\multirow{1}{*}{Graph-Based}
& TAGAT            & 0.92 & 0.93 & 0.91 \\
\midrule
Hybrid
& LLM-GNN Fusion Model & \textbf{0.94} & \textbf{0.95} & \textbf{0.94} \\
\bottomrule
\end{tabular}
\end{table}

\begin{table}[t]
\centering
\caption{Cascade Severity Classification Performance on IEEE-300 bus system using T=4}
\label{class300}
\small
\setlength{\tabcolsep}{3pt}
\begin{tabular}{>{\raggedright\arraybackslash}p{1.85cm}>{\raggedright\arraybackslash}p{2.7cm}ccc}
\toprule
Category & Model & F1 & Precision & Recall \\
\midrule
\multirow{3}{*}{ML}
& SVM-RBF       & 0.66 & 0.67 & 0.64 \\
& Random Forest & 0.79 & 0.79 & 0.79 \\
& DNN           & 0.81 & 0.81 & 0.81 \\
\midrule
\multirow{1}{*}{Transformer}
& FT-Transformer & 0.73 & 0.74 & 0.72 \\
\midrule
\multirow{3}{*}{LLM}
& NTI-LLM     & 0.85 & 0.85 & 0.84 \\
& ATI-LLM & 0.88 & 0.88 & 0.87 \\
& ETI-LLM   & 0.88 & 0.88 & 0.88 \\
\midrule
\multirow{1}{*}{Graph-Based}
 & TAGAT         & 0.90 & 0.90 & 0.90 \\
\midrule
Hybrid
& LLM-GNN Fusion Model & \textbf{0.92} & \textbf{0.92} & \textbf{0.91} \\
\bottomrule
\end{tabular}
\end{table}

\vspace{-0.1cm}
\begin{table*}[h]
    \centering
    \caption{\footnotesize{Performance Evaluation of Cascading Severity Prediction using the proposed TAGAT-LLM Fusion model and two window size $T$  for the IEEE 118 and IEEE 300 Bus System. Note: The presented values are in the format of Precision/ Recall/ F-1 Score.}}
    \resizebox{\textwidth}{!}{%
    \begin{tabular}{l l c c c c c}
        \toprule
        \textbf{Method} & \textbf{Observation Window} & \textbf{Small} & \textbf{Medium} & \textbf{Large} & \textbf{Very Large} & \textbf{Weighted Average}  \\
        \midrule
        \multirow{2}{*}{IEEE 118-Bus System} 
   
        & T=4  &  0.93 / 0.80 / 0.86  & 0.98 / 0.93 / 0.96  &  0.97 / 0.92 / 0.95   &  0.99 / 0.94 / 0.96  &  0.95 / 0.94  / 0.94  \\
        &T=8
 
        & 0.91 / 0.91 / 0.91 & 0.96 / 0.96 / 0.96 & 0.96 / 0.95 / 0.95 & 0.99 / 0.97 / 0.98 & 0.96 / 0.95 / 0.96 \\
        \midrule
        \multirow{2}{*}{IEEE 300-Bus System} 

        & T=4 &  1.00 / 1.00 / 1.00  & 0.97 / 0.95 / 0.96 & 0.89 / 0.85 / 0.87   & 0.82 / 0.79 / 0.80  & 0.92 / 0.91 / 0.92  \\
         & T=8  &  1.00 / 1.00 / 1.00  & 0.98 / 0.98 / 0.98  &  0.94 / 0.90 / 0.92   &  0.92 / 0.90 / 0.91  &  0.95 / 0.94  / 0.95  \\

        \bottomrule
    \end{tabular}
    }
    \vspace{2mm}
    \label{performance evaluation2}
\end{table*}

\vspace{-0.2cm}
\subsection{In-System Performance}

\textcolor{black}{In this subsection, the performance of the proposed TAGAT, three LLM variants, and the Graph-LLM Fusion model is evaluated and compared with that of the baseline ML and transformer models. The training and test data are drawn from the same power system. These evaluations are conducted independently for the IEEE 118-bus and IEEE 300-bus systems.}

\textcolor{black}{All models share a common input of the per-line active power flow for each transmission line $i$, i.e., $(p_i^{(1)}, \ldots,p_i^{(T)})$, 
over the window of the first $T$ cascade steps. 
The cascade window size, \(T\), represents a trade-off between prediction latency and the amount of information available for prediction. The impact of different values of $T$ on the prediction accuracy has been investigated in this work; however, two common window sizes of $T=4$ and $T=8$ are specifically considered for all models to facilitate the performance comparison.
In addition to the common power-flow information, each model utilizes its corresponding set of GSP and non-GSP features, as detailed in the model descriptions in Section IV.}

Tables II and III present the performance comparison for $T=4$. On IEEE-118 dataset, the Fusion model achieves an F1 score of 0.94, matching Random Forest and outperforming TAGAT (0.92) and the standalone LLM variants (0.90–0.93). On IEEE-300, the Fusion model achieves the highest F1 score of 0.92, compared with 0.90 for TAGAT, 0.85–0.88 for the LLM variants, and 0.66–0.81 for conventional baselines.  These results indicate a greater advantage of graph-language fusion on the larger system. Table IV further evaluates the Fusion model using $T=4$ and $T=8$. Increasing the observation window improves its weighted F1 score from 0.94 to 0.96 on IEEE-118 and from 0.92 to 0.95 on IEEE-300. On IEEE-118, the largest improvement occurs in the small cascade class (0.86 to 0.91), while on IEEE-300, the very large cascade class shows a substantial improvement (0.80 to 0.91).

\begin{figure}[t]
    \centering

    \subfloat[IEEE 118-bus system.]{
        \includegraphics[width=3.4in]{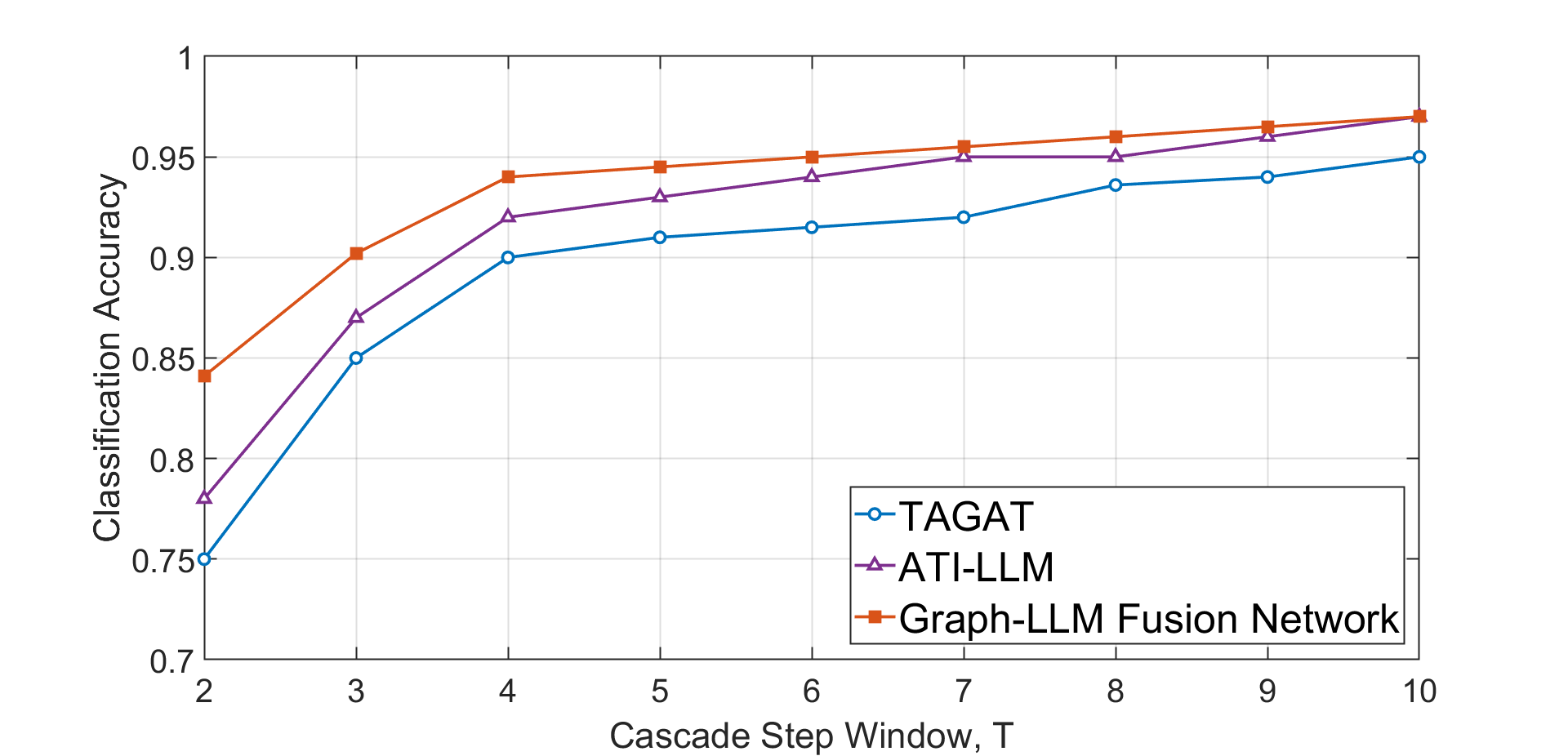}
        \label{fig:accuracy118}
    }

    \vspace{0.0cm}

    \subfloat[IEEE 300-bus system.]{
        \includegraphics[width=3.4in]{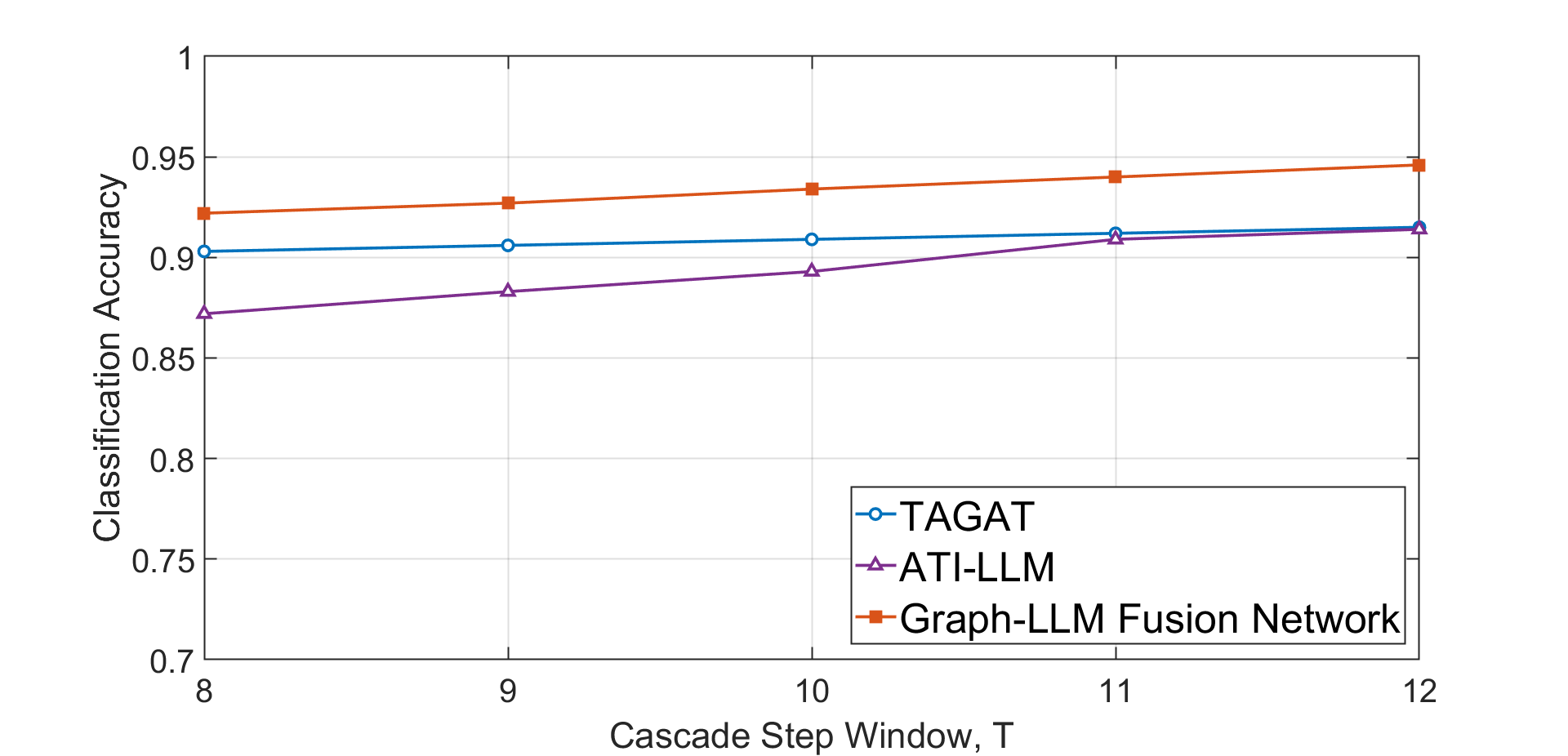}
        \label{fig:accuracy300}
    }

    \caption{\footnotesize{The evolution of classification accuracy as a
function of cascade steps window, $T$ for the standalone TAGAT, ATI-LLM, and Graph-LLM Fusion models:
    (a) IEEE 118-bus system and (b) IEEE 300-bus system.}}
    \label{fig:accuracy_steps}
\end{figure}

Prediction accuracy improves with the number of observed cascade steps, with most gains achieved in the early stages, as shown in Fig.~\ref{fig:accuracy_steps}. For IEEE-118 (Fig.~\ref{fig:accuracy_steps}(a)), the Fusion model consistently achieves the highest accuracy, increasing from 0.84 at Step 2 to 0.94 at Step 4 and approximately 0.97 at Step 10. Similarly, ATI-LLM improves from 0.78 to 0.92 and 0.97, while TAGAT increases from 0.75 to 0.90 and 0.95 at the corresponding steps, narrowing the performance gap as more steps are observed. For IEEE-300 (Fig.~\ref{fig:accuracy_steps}(b)), the fusion model also maintains the highest accuracy, while ATI-LLM and TAGAT gradually converge. This convergence is more pronounced in IEEE-300 because GSP features become available later in the cascade. On average, the first local maximum and uncontrolled islanding are both observed by 7.836 steps for IEEE-118 and 11.416 steps for IEEE-300~\cite{soumav2025learning}. Consequently, the later availability of GSP features in IEEE-300 helps ATI-LLM approach the performance of TAGAT, which does not utilize these features.

\vspace{-0.3cm}

\subsection{Cross-System Transferability Analysis}

To evaluate cross-system transferability, each model is trained on the IEEE 118-bus system and evaluated on the IEEE 300-bus system without further retraining, fine-tuning, or parameter updates. For models employing CORAL, unlabeled target-system data are utilized during training for unsupervised representation alignment, while no target labels are used. This setting evaluates label-free cross-system transfer with unsupervised domain adaptation, followed by few-shot calibration using limited labeled target-system samples. 

Table V summarizes the zero-shot transfer performance. Among the classical baselines, RF achieves the highest accuracy at 0.68, followed by DNN (0.67) and the FT-Transformer (0.65); SVM performs worst among the topology-agnostic baselines at 0.51, and the standalone TAGAT model reaches only 0.56, reflecting its tight coupling to the source-system topology. The standalone LLM attains 0.69 accuracy without topological information, 0.70 with aggregated topological information, and 0.71 with explicit topological information, so each additional block of information adds only a small gain in this setting. The Graph-LLM Fusion model achieves the best zero-shot accuracy overall at 0.72 (0.72 recall and 0.71 precision), indicating that language-based reasoning over structured descriptors transfers more effectively across grid topologies than representations tied to a fixed graph structure.

Table VI presents the few-shot transfer results using 1,000 labeled IEEE-300 samples for target-system calibration. All models improve over their zero-shot performance: SVM, RF, DNN, and FT-Transformer achieve accuracies of 0.56, 0.71, 0.74, and 0.67, respectively. TAGAT shows the largest improvement, from 0.56 to 0.76, outperforming all topology-agnostic models. The standalone LLM achieves 0.71, 0.74, and 0.75 accuracy without, with aggregated, and with explicit topological information, respectively. The Graph-LLM Fusion model achieves the highest accuracy of 0.83 (0.83 recall and 0.82 precision), improving by 0.11 over zero-shot transfer. These results demonstrate that TAGAT benefits most from limited target-system calibration despite its weaker zero-shot generalization, while the Fusion model maintains the strongest cross-system performance in both transfer settings.

\begin{table}[t]
\centering
\caption{Cross-system transferability test: Zero-shot transfer from IEEE-118 to IEEE-300 using T=4.}
\label{tab:zeroshot}
\small
\setlength{\tabcolsep}{4pt}
\begin{tabular}{>{\raggedright\arraybackslash}p{4.2cm}ccc}
\toprule
Model & Accuracy & Recall & Precision \\
\midrule
SVM (RBF Kernel) & 0.51 & 0.52 & 0.51 \\
Random Forest & 0.68 & 0.69 & 0.66 \\
DNN & 0.67 & 0.65 & 0.69 \\
FT-Transformer & 0.65 & 0.66 & 0.64 \\
TAGAT & 0.56 & 0.55 & 0.55 \\
NTI-LLM & 0.69 & 0.69 & 0.69 \\
ATI-LLM & 0.70 & 0.70 & 0.69 \\
ETI-LLM & 0.71 & 0.71 & 0.71 \\
\midrule
Graph-LLM Fusion Model & \textbf{0.72} & \textbf{0.72} & \textbf{0.71} \\
\bottomrule
\end{tabular}
\end{table}

\begin{table}[t]
\centering
\caption{Cross-system transferability test: few-shot transfer from IEEE-118 to IEEE-300 using T=4.}
\label{tab:fewshot}
\small
\setlength{\tabcolsep}{4pt}
\begin{tabular}{>{\raggedright\arraybackslash}p{4.2cm}ccc}
\toprule
Model & Accuracy & Recall & Precision \\
\midrule
SVM (RBF Kernel) & 0.56 & 0.57 & 0.55 \\
Random Forest & 0.71 & 0.72 & 0.71 \\
DNN & 0.74 & 0.74 & 0.73 \\
FT-Transformer & 0.67 & 0.67 & 0.66 \\
TAGAT & 0.76 & 0.77 & 0.75 \\
NTI-LLM & 0.71 & 0.72 & 0.71 \\
ATI-LLM & 0.74 & 0.74 & 0.73 \\
ETI-LLM & 0.75 & 0.74 & 0.76 \\
\midrule
Graph-LLM Fusion Model & \textbf{0.83} & \textbf{0.83} & \textbf{0.82} \\
\bottomrule
\end{tabular}
\end{table}

\vspace{-0.2cm}

\section{Conclusion} 

\textcolor{black}{This work presents an LLM-GNN fusion framework for early-stage cascading failure severity prediction in power transmission networks. The proposed confidence-gated fusion framework combines topology-aware TAGAT representations with the contextual learning capabilities of a LoRA-fine-tuned LLaMA-3-8B model, leveraging complementary structural and system-level information. Evaluations on the IEEE 118-bus and IEEE 300-bus systems demonstrate that the proposed fusion framework achieves competitive or improved predictive performance compared with standalone LLM, graph-based, and conventional machine learning models, highlighting the complementary value of structural and language-based representations. Furthermore, the cross-system transferability analysis highlights the potential of combining graph-based and language-based representations to improve generalization across different network topologies. Future work will focus on extending the proposed framework to predict the exact cascade size and the sequence of transmission line failures, enabling more detailed characterization of cascading events.}
\vspace{-0.2cm}

\section*{Acknowledgment}
This material is based upon work supported by the National Science Foundation under Grant No.~2238658.
\vspace{-0.2cm}

\ifCLASSOPTIONcaptionsoff
  \newpage
\fi

\bibliographystyle{IEEEtran}
\bibliography{IEEEabrv,Bibliography}

\vfill

\end{document}